\documentclass[epsf]{aa}
\usepackage{natbib}
\usepackage{epsfig}

\begin{document}
\title{Accurate $Spitzer$ infrared radius measurement for the hot Neptune GJ 436b
}
\subtitle{}
\author{M. Gillon$^{1,  2}$, B.-O. Demory$^{1}$, T. Barman$^3$, X. Bonfils$^4$, T. Mazeh$^5$, F. Pont$^1$, S. Udry$^1$, M. Mayor$^1$, D. Queloz$^1$}    

\offprints{michael.gillon@obs.unige.ch}
\institute{$^1$  Observatoire de l'Universit\'e de Gen\`eve, 1290 Sauverny, Switzerland\\
$^2$ Institut d'Astrophysique et de G\'eophysique,  Universit\'e de Li\`ege,  4000 Li\`ege, Belgium \\ 
$^3$ Lowell Observatory, 1400 West Mars Hill Road, Flagstaff, AZ 86001, USA\\
$^4$ Observat\'orio Astron\'omico de Lisboa, Tapada da Ajuda, P-1349-018 Lisboa, Portugal\\
$^5$ School of Physics and Astronomy, Raymond and Beverly Sackler Faculty of Exact Sciences, Tel Aviv University, Tel Aviv, Israel\\  
}	

\date{Received date / accepted date}
\authorrunning{M. Gillon et al.}
\titlerunning{Accurate $Spitzer$ radius for the hot Neptune GJ 436b}
\abstract{We present $Spitzer$ $Space$ $Telescope$ infrared photometry of a primary transit of the hot Neptune GJ 436b. The observations were obtained using the 8 $\mu$m band of the InfraRed Array Camera (IRAC). The high accuracy of the transit data and the weak limb-darkening in the 8 $\mu$m IRAC band allow us to derive (assuming $M$ = 0.44 $\pm$ 0.04  $M_\odot$ for the primary) a precise value for the planetary radius (4.19$^{+0.21}_{-0.16}$  $R_\oplus$), the stellar radius (0.463 $^{+0.022}_{-0.017}$ $R_\odot$), the orbital inclination (85.90$^\circ$$^{+0.19^\circ}_{-0.18^\circ}$) and transit timing (2454280.78186 $^{+0.00015}_{-0.00008}$ HJD).  Assuming current planet models, an internal structure similar to that of Neptune with a small H/He envelope is necessary to account for the measured radius of GJ 436b. 
 }

\maketitle

\section{Introduction}

While more than two hundred planets have been detected outside our Solar System, the minority of them that transit a bright parent star have had the highest impact on our overall understanding of these objects (see review by Charbonneau et al. 2007). Indeed, the structure and atmospheric composition of the non-transiting extrasolar planets detected by radial velocity (RV) measurements remain unknown because the only information available from the RV time series are the orbital parameters (except the inclination of the orbital plane and the longitude of the ascending node), and only a lower limit for the planetary mass. Transiting planets are the only ones for which  accurate estimates of mass, radius, and, by inference, composition can be obtained. The brightest of these systems can be monitored during primary and secondary transits with high-precision instruments, allowing us to characterize their composition and atmosphere, and learn what these other worlds look like. 

Since 2005, many exciting results on transiting planets have been obtained with the  $Spitzer$ $Space$ $Telescope$: the detection of the thermal emission from four giant planets (Charbonneau et al. 2005; Deming et al. 2005, 2006; Harrington et al. 2007), the precise measurement of an infrared planetary radius (Richardson et al. 2006), the measurement of the infrared spectrum from two planets (Richardson et al. 2007; Grillmair et al. 2007), and the measurement of the phase-dependent brightness variations from HD 189733b (Knutson et al. 2007). These results have demonstrated the very high potential of the $Spitzer$ $Space$ $Telescope$ to characterize transiting planets and motivated many theoretical works (e.g. Barman et al. 2005; Burrows et al. 2005; Williams et al. 2006; Barman 2007).

All of these studies have targeted gaseous giant planets. Many Neptune-mass and even Earth-class planets have been detected by the RV technique (e.g. Rivera et al. 2006; Bonfils et al. 2007; Udry et al. 2007), but until very recently none of these small mass planets had been caught in transit. Remarkably, the transits of a known Neptune-sized planet have just been detected (Gillon et al. 2007, hereafter G07). The host star, GJ 436, is a very close-by M-dwarf ($d$ = 10.2 pc). The planet itself was first detected by RV measurements (Butler et al. 2004, hereafter B04; Maness et al. 2007, hereafter M07). It is by far the closest, smallest and least massive transiting planet detected so far. Its mass is slightly larger than Neptune's at $M$ = 22.6 $\pm$ 1.9 $M_\oplus $, while its orbital period is 2.64385 $\pm$ 0.00009  days (M07). The shape and depth of the ground-based transit lightcurves show that the planet is crossing the host star disc near its limb (G07).

Assuming a stellar radius $R$  = 0.44 $\pm$ 0.04 $R_\odot$, G07 measured a planet size comparable to that of Uranus and Neptune. Considering this measurement and current planet models,  GJ 436b should be composed of refractory species, with a large fraction of water ice, probably surrounded by a thin H/He envelope. Nevertheless, the original uncertainty on the radius presented in G07 is too large ($\sim 10 \%$) to confidently claim the presence of the H/He envelope. Indeed, the lower end of the G07 1-sigma radius range is close to the theoretical mass-radius line for a pure water ice planet from Fortney et al. (2007). Consequently, GJ 436b could be an ice giant planet very similar to Neptune or an "ocean planet" (L\'eger et al. 2004), and a more precise radius measurement is needed to discriminate between these two compositions.

The planetary radius uncertainty in G07 was mainly due to the error on the radius of the primary. Principally because of the presence of correlated noise (Pont et al. 2006), ground-based photometry of a shallow transit like the one of GJ 436b is generally not accurate enough to break the degeneracy between the impact parameter and the primary radius, and fails to allow an independent determination of the primary radius. A prior constraint on the stellar radius based on evolution models and/or observational constraints is needed, limiting the final precision on the planetary radius. This is no more the case with high signal-to-noise space-based photometry. In this case, both radii can be determined very precisely from the lightcurve analysis, assuming only the primary mass or a stellar mass-radius relation. In the case of GJ 436b, the $Spitzer$ telescope is particulary well suited for this task. Indeed, the host star is rather bright in the near-infrared (K $\sim$ 6), and the weak limb-darkening at these wavelengths is a huge advantage, as demonstrated by the recent HD 209458b and HD 189733b radius measurements (Richardson et al. 2006; Knutson et a. 2007). 

We report here the $Spitzer$ observations of a primary transit of GJ 436b within the 8 $\mu$m band of the InfraRed Array Camera (IRAC; Fazio et al. 2004). The analysis of the observations allows us to obtain a  precise measurement of the primary and planetary radii, bringing an important constraint on the composition of the planet, which reveals to be very similar to Neptune. 

Section 2 describes the observations, the reduction procedure and the resulting photometry.  Our analysis of the obtained time series is described in Section 3. Our conclusions are presented in Section 4.

\section{Observations and data reduction}

GJ 436 was observed on June 29th UT for 3.4h covering the transit, resulting in 28480 frames. On top of that, a blank field located a few minutes away from the target was imaged just before and after the observations in order to characterize pixel responses and to improve overall field flatness.

Data acquisition was made using IRAC in its 8$\mu$m band, subarray mode, with an effective exposure time of  0.32s to avoid saturation problems due to the high brightness of GJ 436 at these wavelengths. Observing  with IRAC in one channel only avoids repointing processes which reduces observational time efficiency. Subarray mode consists of windowing the full array into a 32x32 pixels window, allowing a better time sampling. Each resulting data file consists of a set of 64 of these windowed images, delivered to the community as BCD (Basic Calibrated Datasets) files after having been processed by a dedicated pipeline at Spitzer Science Center. We combine each set of 64 images using a 3-sigma clipping to get rid of transient events in the pixel grid, yielding 445 stacked images with a  temporal sampling of $\sim$ 28s. 

Using the time stamps in the image headers, we compute the JD corresponding to the center of each integration, and add the relevant heliocentric correction to obtain the date in Heliocentric Julian Date (HJD). We convert fluxes from the Spitzer units of specific intensity (MJy/sr) to photon counts, and aperture photometry is performed on the stacked images. The aperture radius giving the best compromise between the noise from the sky background and from the centroid variations is found to be 4.0 pixels. An estimate for the sky background is derived from an annulus of 12--24 pixels and subtracted from the measured flux. 

As already noticed by similar works (e.g. Charbonneau et al. 2005), time series obtained with IRAC in the 8$\mu$m band show a gradual detector-induced rise of the measured signal, probably due to charge trapping. Despite this instrumental effect, the eclipse can be clearly seen in the raw data (see Fig. 1). To correct for the instrumental rise effect, we zero-weight the eclipse and the 80 first points of the time series (to avoid the steepest part of the rise in our modeling of the effect) and we divide the lightcurve by the best fitting asymptotic function with three free parameters. We then evaluate the average flux outside the eclipses and use it to normalize the time series\footnote{Our final photometric time series is available only in electronic form at the CDS via anonymous ftp to cdsarc.u-strasbg.fr (130.79.128.5) or via http://cdsweb.u-strabg.fr/cgi-bin/qcat?J/A+A/}. The $rms$ of the resulting lightcurve evaluated outside the eclipse is 0.7 mmag.
\begin{figure}
\label{fig:a}
\centering                     
\includegraphics[width=9.0cm]{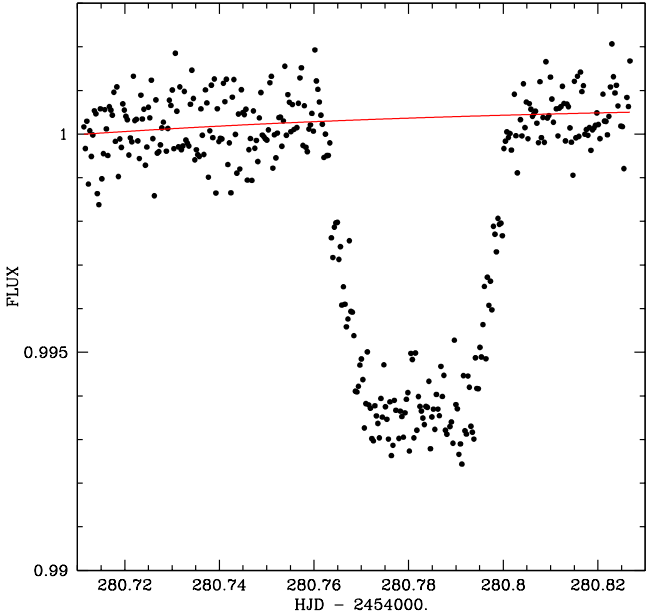}
\caption{Raw IRAC 8$\mu$m photometric time series for the primary transit of GJ 436b (arbitrary unit).  The best fitting asymptotic function for the instrumental rise effect is superimposed (solid line).}
\end{figure}

\section{Time series analysis}

Despite its closeness to its host star, GJ 436b exhibits a clear orbital eccentricity $e \sim 0.16$ (B04; M07). To properly analyze the primary transit, the eccentricity and the argument of the periapse have to be taken into account. In the case of a circular orbit, the formula connecting the projected separation of the centres (in units of the primary radius) $z$ and the time of observation is: 

\begin{equation}\label{eq:a}
z = \frac{a}{R_\ast} [(\sin n t)^2 + (\cos i \cos n t)^2] ^{1/2}\textrm{,}
\end{equation}

where $a$ is the semi-major axis, $R_\ast$ is the stellar radius, $n$ is the mean motion $2\pi / P$, and $i$ is the orbital inclination. In case of a non-zero orbital eccentricity $e$, equation (1) becomes:

\begin{equation}\label{eq:a}
z = \frac{r}{R_\ast} [(\sin n_2 t)^2 + (\cos i \cos n_2 t)^2] ^{1/2}\textrm{,}
\end{equation}

where $r$ is the orbital distance and $n_2$ the angular frequency at the orbital location of the transit. $r$ and $n_2$ are given by the formulae:

\begin{equation}\label{eq:b}
r_\ast =  \frac{a (1 - e^2)}{(1 + e \cos f)}\textrm{,}
\end{equation}

\begin{equation}\label{eq:c}
n_2 =  n \frac{(1 + e \cos f)^2}{(1 - e^2)^{3/2}}\textrm{,}
\end{equation}

where $f$ is the true anomaly, i.e. the angle between the transit location and the periapse. The formula connecting this latter and the argument of periapse $\omega$ is simply:

\begin{equation}\label{eq:c}
f = \frac{\pi}{2}  - \omega\textrm{.}
\end{equation}

Using the values for $e$ and $\omega$ from M07 and equation  (2), transit profiles are fitted to the primary transit data using the Mandel \& Agol (2002) algorithm, the orbital elements in M07, and quadratic limb darkening coefficients\footnote{$a = 0.045 \pm 0.01$ and $b = 0.095 \pm 0.015$}  derived from a stellar model atmosphere with $T_{eff}$ = 3500 K, $\log g$ = 4.5 and [Fe/H] = 0.0. As in G07, the mass of the star is fixed to $M = 0.44 \pm 0.04$ $M_\odot$.  

The free parameters of the fit are the radius of the planet $R_p$ and the primary $R_\ast$, the orbital inclination $i$ and the central epoch of the transit $T_p$. Starting from initial guess values for these parameters, the $\chi^2$ of the fit is computed over a large grid of values. The grid cell corresponding to the lowest $\chi^2$ is then considered as the starting point of the next step. The same process is then repeated twice, using a finer grid. At this stage, the downhill-simplex AMOEBA algorithm (Press et al. 1992) is used to reach the $\chi^2$ minimum. Figure 2 shows the best fitting theoretical curve superposed on the data. 

To obtain realistic error bars on the free parameters, we have to take into account the uncertainty coming from the stellar and orbital parameters, and also the one coming from the initial guess for the free parameters. The possible presence of correlated noise in the lightcurve has also to be considered (Pont et al. 2006). In particular, it is clear that the calibration of the detector-induced rise of the signal is able to produce systematic errors. 

To account for these effects, we use the following Monte-Carlo procedure. A high number (10,000) of lightcurve fits is performed. For each fit, we use randomly chosen orbital and stellar parameters from a normal distribution with a width equals to the published uncertainties and randomly chosen initial guess values for the free parameters from a wide rectangular distribution. Furthermore, the residuals of the initial fit are shifted sequentially about a random number, and then added to the eclipse model. The purpose of this procedure called the ``prayer bead'' (Moutou et al. 2004) is to take into account the actual covariant noise level of the lightcurve. We estimated the error bars from the distribution of the 10,000 derived values for the free parameters. The obtained value for $R_p$, $R_\ast$, $i$ and $T_p$ and their error bars are given in Table 1.  

To test the influence of the eccentricity on the obtained values for the radii and the orbital inclination, we made a fit assuming a circular orbit, obtaining $R_p = 4.15$ $R_\oplus$, $R_\ast = 0.459$ $R_\odot$, $i = 85.97^o$ (impact parameter = 0.842), and $T_p = 2454280.78188 $ HJD. The influence of the eccentricity reveals thus to be weak. This is due to the fact that $f$ at transit is close to 90$^o$, i.e., the transit location is nearly perpendicular to the periapse.  

\begin{figure}
\label{fig:c}
\centering                     
\includegraphics[width=9.0cm]{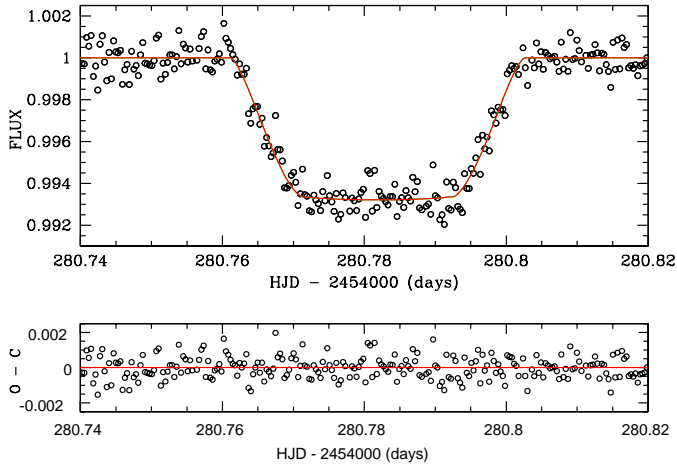}
\caption{$Top$: Final time series for the primary transit. The best-fit theoretical curve is superimposed.  $Bottom$: The residuals of the fit ($rms$ =  0.7 mmag).}
\end{figure}

While fixing the primary mass in the fit is justified by the fact that spectroscopic and photometric observations bring a strong constraint on this parameter (see discussion in M07), it does not benefit from the fact that the shape of the transit is not governed by the primary radius $R_\ast$ but by the density of the primary $\rho_\ast$ (Seager \& Mall\'en-Ornelas, 2003). Instead of assuming a primary mass but a stellar mass--radius relation,  one can determine $\rho_\ast$, $R_p/R_\ast$, the impact parameter $b$ and the transit timing $T_p$, then use the assumed mass-radius relation to obtain $R_\ast$, and finally get $R_p$ from the measured radii ratio. To test the influence of the used prior constraint on the host star (mass $vs$ mass--radius relation), we perform a new fit to the data considering $\rho_\ast$, $R_p/R_\ast$, the impact parameter $b$ and $T_p$ as free parameters, and used the relation $M_\ast = R_\ast$ (see Ribas et al. 2006 and references therein) to ultimately determine the stellar and planetary radius. At the end, we obtain $R_p = 4.31$ $R_\oplus$, $R_\ast = 0.477$ $R_\odot$, $i = 85.84^o$ (impact parameter = 0.859), and $T_p = 2454280.78190 $ HJD. These values are within the 1-sigma error bars of our fit assuming a fixed value for the stellar mass. This agreement between two independent fits using a different prior constraint on the host star  is a good indication of the robustness of our  solution presented in Table 1. 

\begin{table}
\begin{tabular}{l l } \hline\hline
\\
Stellar Radius [$R_\odot$] &  $0.463^{+0.022}_{-0.017}$ \\[4pt]
Orbital inclination [$^\circ$] & $85.90^{+0.19}_{-0.18}$ \\[4pt]
Impact parameter &  $0.849^{+0.010}_{-0.013}$ \\[4pt]
Planet Radius [$R_\oplus$]&  $4.19^{+0.21}_{-0.16}$ \\[4pt]
\ \ \ \ \ \ \ \ \ \ \ \ \ \ \ \ \ \ \ \ [km]&  $26720^{+1340}_{-1020}$\\[4pt]
Mid-transit timing [HJD] & $2454280.78186^{+0.00015}_{-0.00008}$\\[4pt]
 \hline\hline 
 & \\
\end{tabular}
\caption{Parameters derived in this work for the GJ 436 system, host star and transiting planet.} 
\label{param}
\end{table}

\section{Conclusions}

The exquisite quality of the $Spitzer$ photometry allows the breaking of the degeneracy between the impact parameter and the primary radius, leading to a significantly more accurate radius measurement than the one presented in G07. Comparing our new value for the planetary radius to the theoretical models of Fortney et al. (2007), and using $M = 22.6 \pm 1.9 $ $M_\oplus $, we notice that GJ 436b is now more than 3-sigmas away from the "pure ice" composition mass-radius line (see Fig. 3). 

As noticed in G07, the presence of a significant amount of methane and ammonia in addition to water  within a pure ice planet could slightly increase the radius above the theoretical value for a pure water ice planet. Nevertheless, a planet composed only of ice (and thus without any rock) and widely enriched in ammonia and methane is largely improbable in the current paradigm: all the icy objects in the Solar System have a considerable fraction of rock and have their ice composition dominated by water. The presence of an H/He envelope can thus be considered as very likely. The first transiting hot Neptune appears to be very similar to our Neptune. 

Before the detection of the transits, Lecavelier des Etangs (2007) showed theoretically that GJ 436b could not be a low mass gaseous planet. Indeed, a density lower than $\sim$ 0.7 g cm$^{-3}$ would have led to an intense atmospheric evaporation. The density deduced from our radius measurement ($\sim $ 1.7 g cm$^{-3}$) is well above this limit. Despite the closeness to the host star, the H/He envelope of the planet should thus be stable and resistant to evaporation over long timescales. 

\begin{figure}
\label{fig:d}
\centering                     
\includegraphics[width=9.0cm]{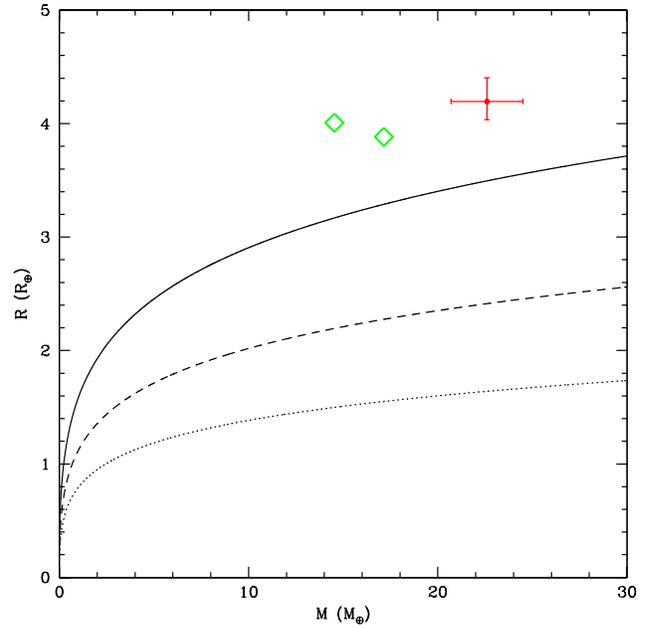}
\caption{Location of GJ 436b in a planetary mass-radius diagram, compared to the one of Uranus and Neptune (open diamonds) and the theoretical mass-radius relations from Fortney et al.  (2007) for pure water ice (solid line), pure rock (dashed line) and pure iron (dotted line) planets. }
\end{figure}
 
\begin{acknowledgements} 
This work is based on observations  made with the $Spitzer$ $Space$ $Telescope$, which is operated by the Jet Propulsion Laboratory, California Institute of Technology, under NASA contract 1407. X.B. acknowledges support from the Funda\c{c}\~ao para a Ci\^encia e a Tecnologia (Portugal) in the form of a fellowship (references SFRH/BPD/21710/2005). B.-O.D. acknowledges the support of the Fonds National Suisse de la Recherche Scientifique. G. Tinetti and O. Grasset are greatfully acknowledged for providing us informations about icy objects in the Solar System. We are also grateful to P. Magain and D. S\'egransan for helpful discussions, and to F. Allard for her support and suggestions. We thank A. Ofir for his contribution. We acknowledge the referee I. Snellen for the fast review of the paper and valuable suggestions. 
\end{acknowledgements} 

\bibliographystyle{aa}
{}
\end{document}